# How Many Simulations Do We Exist In? A Practical Mathematical Solution to the Simulation Argument

Hutan Ashrafian

*Institute of Global Health Innovation, Imperial College London and Institute of Civilisation*

**Correspondence to:**

Dr Hutan Ashrafian

Institute of Global Health Innovation,

Imperial College London,

10th Floor Queen Elizabeth the Queen Mother (QEQM) Building,

St Mary's Hospital,

Praed Street, London,

W2 1NY, UK

h.ashrafian@imperial.ac.uk

**Abstract**

The Simulation Argument has gained significant traction in the public arena. It has offered a hypothesis based on probabilistic analysis of its assumptions that we are likely to exist within a computer simulation. This has been derived from factors including the prediction of computing power, human existence, extinction and population dynamics, and suggests a very large value for the number of possible simulations within which we may exist. On evaluating this argument through the application of tangible 'real-world' evidence and projections, it is possible to calculate real numerical solutions for the Simulation Argument. This reveals a much smaller number of possible simulations within which we may exist, and offers a novel practicable approach in which to appraise the variety and multitude of conjectures and theories associated with the Simulation Hypothesis.

**Introduction**

The Simulation Argument (SA) has gained significant public momentum and prominence in the contemporary public zeitgeist. This is because it exquisitely captures the large range of possibilities offered by the range, depth and fidelity of digital technology to address fundamental existential questions of humanity. Whilst not necessarily intended to be applied in a multitude of settings, it has served to support a vast range of arguments that range from extensions of the philosophical brain-in-a-vat thought experiment to modern rationalizations of religions, explanation of astronomical dynamics and multiple universe theories. It assumes that (i) advanced computer and advanced computability will allow the existence of (artificial) intelligent agents that can pass the Turing Test or its future modifications [1], and (ii) that these can be generated on extensive simulations which can accommodate multiple intelligent agents. At its core, the SA [2] has three core tenets:

(1) The human species is very likely to go extinct before reaching a "post-human" stage
(2) Any post-human civilization is extremely unlikely to run a significant number of simulations of their evolutionary history (or variations thereof)
(3) We are almost certainly living in a computer simulation

The last of these three, also known as the Simulation Hypothesis (SH), has concurrently received the most acclaim and controversy by suggesting that "we are almost certainly living in a computer simulation." This argument and its hypothesis have been the subject of several points of contention [3] that have subsequently resulted in clarifications ranging from issues of epistemological externalism and the highlighting of difference from the traditional brain-in-a-vat argument [4].

More recently, the originator of this argument has also modified the mathematical assumptions and construct of the Simulation Argument by adding either of two conceptual mathematical 'patches' [5] to address and correct for perceived vulnerabilities. Constraints have also been suggested regarding the mathematical plausibility of large-scale universe simulations based on the extrapolation of the hypothesis [6]. The patches include:

(4a) the average number of individuals living in a pre-posthuman phase is not astronomically different in civilizations who do not run simulations compared to those who run multitudes of simulations (as a significant number of *unusually brief pre-post-human phases* may result in many post-human observers living outside simulations.)
Or
(4b) Even if pre-posthuman civilizations not running ancestor simulations had longer phases of time where they could have exposure to human experiences, then these civilizations would not contain (on average) more people with exposure to human experiences than those civilizations running ancestor simulations.

Many of the arguments against the original hypothesis are founded on precepts of a hyper-advanced post-human civilization in the distant future, simulating the history of the universe and the civilization's own 'ancestors' within it. As a result, the mathematical and philosophical assumptions there-in can become warped due to possibilities of extreme variance in prediction at a distant future. My aim is to offer a pragmatic approach based on current perceptible 'real-world' evidence to identify some more tangible understandings regarding the original Simulation Argument and to solve it to calculate the number of simulations that we exist within.

**The Post-Human Stage**

A post-human civilization can also be defined as an advancement of the level of civilization to a higher order. One measure of this has been previously defined by the Soviet astronomer Nikolai Kardashev [7] who has established a scale of civilization based on its capacity to generate and apply energy (Type I – Planetary, Type II – Stellar, Type III – Galactic). According to Sagan's contemporary modification of this scale[8, 9] we are currently a civilization scoring 0.72 and we will become a Type I civilization when we can utilize all the energy that reaches a planet from its parent star (corresponding to $10^{16}$-$10^{17}$ watts) [10]. It has been predicted that we will achieve this post-human stage of a Type I Kardashev civilization by 2100 [11].

To generate a realistic human simulation, approximately $10^{33}$-$10^{36}$ operations was originally estimated as a rough estimate [2]. Based on current trends at the beginning of the 3rd decade of the 21st century, we have achieved $10^4$ MegaFLOPS (Floating Point Operations Per Second) per watt or $10^{10}$ FLOPS per watt in our most efficient computers [12] (and therefore beyond $10^{10}$ operations per second). We have been improving by $10^{0.2}$ MegaFLOPS a year, so that at the dawn of the post-human stage in 2100 we may increase our operation processing ability by approximately $10^{16}$ in 80 years amounting to $10^{26}$ operations per second per watt. Adopting the possibility of a

Type I Kardashev status by 2100 ($10^{16}$-$10^{17}$ watts), this would correspond to a computing processing capability of to $10^{42}$-$10^{43}$ operations per second. This processing power could feasibly address the computational needs of an 'ancestor simulation'. Furthermore, even if this trend tails off (which remains a possibility), the potential for Quantum computers is also expanding, such that current Quantum Computing simulations [13], can achieve $10^{17}$ FLOPS at $10^{9}$ FLOPS/Watt (better than current non-Quantum supercomputer efficiency) and this technology with the prospect of Quantum Supremacy is likely offer even more performance at higher power efficiency for the potential to complete ancestor simulations by 2100.

**Methods**

The Mathematical Formula of the Original Simulation Argument and its Patches

The original Simulation Argument [2] consisted of the following notation:

*P(DOOM)*: The probability that humanity goes extinct before reaching the post-human stage

*1-P(DOOM)*: The probability that humanity survives extinct before reaching the post-human stage

$\overline{N}$: Average number of ancestor-simulations run by a post-human civilization

$\overline{H}$: Average number of individuals that have lived in a civilization before it reaches a posthuman stage

*f$_{sim}$*: is the fraction all observers with human-type experiences that live in simulations

$f_I$: is the fraction of post-human civilizations that are interested *and technically able* in running ancestor-simulations

$\overline{N_I}$: Average number of post-human civilizations that run ancestor-simulations

*P(SIM)*: is the probability that you are in a simulation.

*s*: Total number of (all) post-human civilizations that run ancestor-simulations

*n*: Total number of (all) post-human civilizations that do not run ancestor-simulations

*N*: Total number (large) of ancestor-simulations run by post-human civilizations (*s*)

$H_n$: The average number of pre-posthuman beings in *n* civilizations that do not run at least *N* simulations

$H_s$: The average number of pre-posthuman beings in *n* civilizations that do not run at least *N* simulations

The original Simulation Argument equation was therefore:

$$P(SIM) = f_{sim} = \frac{[1 - P(DOOM)] \times \overline{N} \times \overline{H}}{([1 - P(DOOM)] \times \overline{N} \times \overline{H}) + \overline{H}}$$

Consequently, the following was proposed: (1) *P(DOOM)* ≈*1*, (2) *f$_I$* ≈ *0*, (3) *P(SIM)* ≈ *1*.

At the root of the original argument, it was argued that:

$$\overline{N} \geq f_I \overline{N_I}$$

And furthermore $\overline{N_I}$ (the average number of post-human civilizations that run ancestor-simulations) is 'extremely large', so that $\overline{N} \times \overline{H}$ can be substituted by $f_I \overline{N_I}$ to calculate the probability of being in a simulation *P(SIM)*.

I argue that $\overline{N_I}$ does not necessarily have to be extremely large as this would not inevitably alter the original argument or its equation dynamics, or that that $\overline{N} \times \overline{H}$ has to be substituted by $f_I \overline{N_I}$. Furthermore, I suggest that the original formula (above) could be applied to 'tangible' 'real-world' data in order to calculate the probability of being in a simulation *P(SIM)*.

Although the author of the original SA did not formally describe it as a hierarchical argument, I consider that it fits well into this category (**Figure 1**).

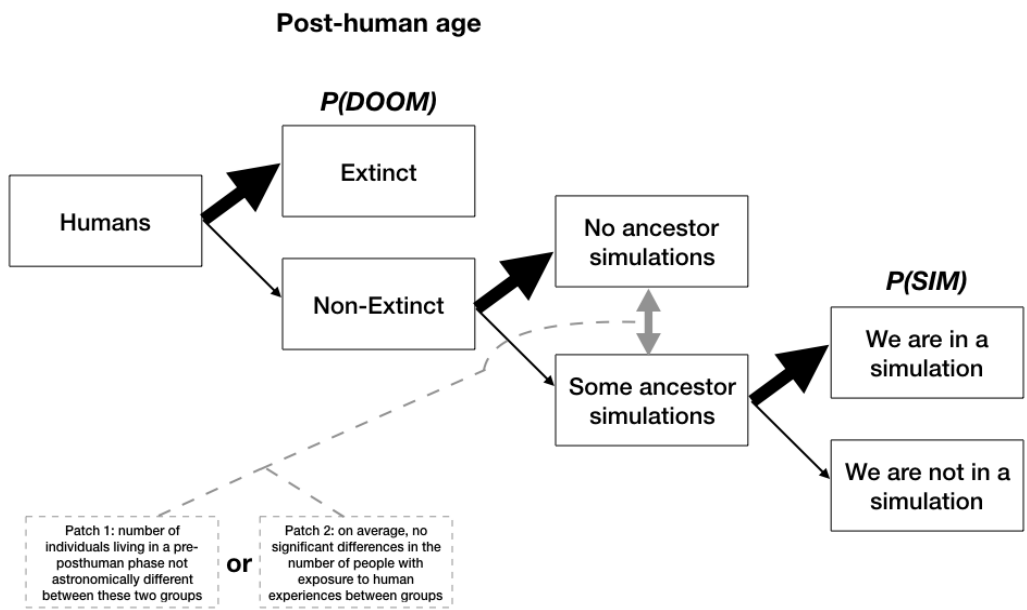

**Figure 1.** Hierarchical structure of the Simulation Argument

The vulnerability patch by SA's author [5] was offered after the finding that if there were a significant number of *unusually brief pre-post-human phases*, this may result in many post-human observers living outside simulations. The patch addressed this with the 'weak' assumption that "the typical duration (or more precisely, the typical cumulative population) of the pre-post-human phase does not differ by an astronomically large factor between civilizations that never run a significant number of ancestor simulations and those that eventually do", so that:

$$H_n/H_s \leq N/1000000$$

And

The fraction of civilizations that never reach the post human phase is $\geq 99\%$

This will also hold true for calculations here-in, where there is a concurrent assumption that unusually brief pre-post-human phases are by definition probabilistically unusual.

The second patch would also apply, where the exposure of human experiences to individuals within populations would on average be comparable in quantity in a civilization with many ancestor simulations to civilizations without ancestor simulations.

With these assumptions I utilize the SA equation to calculate: $\overline{N}$ (average number of ancestor simulations in a post-human civilization – this could represent our civilization) and *P(SIM)* (probability of living in a simulation) for different values of $\overline{H}$ (average number of individuals that have lived before reaching a post-human stage) and *P(DOOM)* (probability of extinction for that civilization).

**Estimating the Values for Equation Variables**

Continuing the assumption that a post-human phase corresponds to a Type I Kardashev civilization that can be reached by 2100, the following can be determined:

*P(DOOM)*: 5% and *1-P(DOOM)*: 95% based on the prediction [14] of global annihilation projected from multiple factors ranging from nuclear war, global pandemics, unknown consequences and other risks that may reach an infinite threshold of world destruction.

$\overline{H}$: Approximately 120 billion cumulative individuals up to the year 2100 based on the area under the curve following the finding that from commencement until the year 2010 the cumulative world population was 108 billion (100.8 x$10^9$) [15] and the projected population in 2100 [16, 17] will be 10.875 billion people.

**Results**

Applying these values, we can calculate the values of *P(SIM)* at different values of $\overline{N}$. This reveals:

There is a 90% chance that we are in one of 10 simulations (advanced enough to accommodate our complexity)

There is a 95% chance that we are in one of 20 simulations (advanced enough to accommodate our complexity)

There is a 99% chance that we are in one of 100 simulations (advanced enough to accommodate our complexity)

There is a 99.9% chance that we are in one of 701 simulations (advanced enough to accommodate our complexity)

There is a 99.99% chance that we are in one of 2105 simulations (advanced enough to accommodate our complexity)

Partial differentiation with respect to $\overline{H}$, demonstrated a low level of sensitivity. As a result, further calculations of *P(SIM)* increasing the value of $\overline{H}$ (average number of individuals before reaching a post-human stage) by 100 orders of magnitude and increasing the likelihood of world extinction from 5 to 10% (**Table 1**) did not cause substantial changes in these results and suggested the robustness of the solved outcome.

| P(SIM) | P(DOOM)=5% $H=1.2\times10^{11}$ | P(DOOM)=5% $H=1.2\times10^{111}$ | P(DOOM)=10% $H=1.2\times10^{11}$ | P(DOOM)=10% $H=1.2\times10^{111}$ |
|---|---|---|---|---|
| 90% | 10 | 10 | 10 | 10 |
| 95% | 20 | 20 | 21 | 21 |
| 99% | 100 | 100 | 105 | 105 |
| 99.9% | 701 | 701 | 740 | 740 |
| 99.99% | 2105 | 2105 | 2222 | 2222 |

**Table 1**. Calculation of $\overline{N}$ (average number of ancestor simulations), at different probabilities of *P(SIM)* (probability of living in a simulation) for different values of $\overline{H}$ (average number of individuals that have lived before reaching a post-human stage) and *P(DOOM)* (probability of extinction).

**Discussion**

This analysis reveals a numerical solution to the Simulation Argument and offers a new and deeper comprehension of the dynamics of the Simulation Hypothesis based on the tangible projection of world events.

The identification that there is a 99% probability that our existence lies within an average of 100 simulations and a 99.9% probability that our existence lies within an average of 701 simulations is a novel finding. Most Simulation Hypothesis advocates (and detractors) typically imply or assume almost an immeasurable number of simulations taking place in the context of a post-human civilization, and that we exist in one of a very large number of simulations.

These results suggest a new finding, that rather than being in one of countless simulations, if the Simulation Argument holds true, then we may exist in one of a handful of simulated worlds. This adds a layer of plausibility in the original Simulation Hypothesis, where it would be much more feasible to exist in a few high-fidelity simulations that could mimic high degrees of complexity apposite with our current civilization.

Since the era of Turing, it has been widely proposed that given enough computing resources and time, simulated evolution could be performed to generate simulated universes, worlds and humans. The situation of countless simulations of high fidelity would need exhaustive energy and processing power and whilst theoretically possible may not represent a practical reality. Furthermore, the author of the original SA suggested the added theoretical possibility of state of nested simulations hierarchically existing within each other. Whilst the increase in the number of possible simulations increases the chances of living in a simulation, this would not account for the likely loss of fidelity of a simulation-within-a-simulation.

Using this practical numerical approach, increasing the value of $\overline{H}$ (average number of individuals before reaching a post-human stage) by several orders of magnitude does not noticeably alter the number of simulations that corresponds to the probability of being in a simulation (**Table 1**).

This analysis also offers a novel 'upstream' angle to the Simulation Argument. Here rather than assuming that we are *necessarily* in a 'downstream' simulation where post-human civilizations are simulating their ancestors, the low number of simulations needed to achieve our possible simulated existence may allow the possibility of a less advanced civilization to simulate us (a more advanced civilization) 'upstream' purely of the probability function of the infinite monkey theorem. Here the sheer chance of simulating a more advanced simulation could be more numerically possible if a lesser advanced civilization built enough heuristic simulation platforms that by chance would develop superior beings (our civilization) within it.

The implication that we might be in only a handful of simulations carries some novel inferences for the Simulation Hypothesis. Firstly the prevailing attitude associated with the hypothesis that our whole existence is based on the sheer multitude of simulations associated with multiple universes or *multiverses* does not necessarily need to exist as an *a priori*. Secondly the associated presupposition that our specific existence is based on an evolutionary-type model where the massive number of simulations may lead to a rare '*goldilocks*' environment that would simulate us in our exact current format is also not necessarily securable. As such, comparisons with the Drake Equation [18] representing the uniqueness of our existence based on numerical probability seem unjustifiable for this simulation setting. Consequently, a third concept of existing with such a small number of simulations renders much more resonance with a targeted simulation by design.

This numerical approach also generates a multitude if its own questions, if the assumptions hold true, then exactly which number of simulation are we? Are we the first or last of these handful of simulations? Now that we have a finite number with which to associate ourselves, are we more or less unique than we understood before? Are the numbers for such an equation those '*forced*' on us through the simulation and can we even extrapolate beyond any simulation within which we are the simulatees?

The results from this analysis remain quite robust despite increasing the likely population of humans and by decreasing the likelihood of human species survival. Suggesting that if the assumptions and equations hold true. The results herein remain reliable.

**Conclusions**

A numerical solution to the Simulation Argument and its associated hypothesis that we exist in a simulation has offered a new angle through which to consider such a possibility. This reveals a high probability of existence within only a handful of simulations. This overturns the implied 'reality' or current orthodoxy of massive numbers of simulations taking place in an era of immense computing capacity. Rather it offers the novel vantage that even with colossal processing power and speed, our possible existence within a simulation may derive from a much more targeted simulation with underlying precision. This practical approach that applies real-world projections may help appraise the variety and multitude of conjectures and theories associated with the Simulation Hypothesis and can contribute to the dialogue of our deeper existence.

# References

[1] Ashrafian H, Darzi A, Athanasiou T. A novel modification of the Turing Test for artificial intelligence and robotics in healthcare. Int J Med Robot 2015;11:38-43.
[2] Bostrom N. Are we living in a computer simulation? Philos Quart 2003;53:243-255.
[3] Weatherson B. Are you a Sim? Philos Quart 2003;53:425-431.
[4] Bostrom N. The simulation argument: Reply to Weatherson. Philos Quart 2005;55:90-97.
[5] Bostrom N, Kulczycki M. A patch for the Simulation Argument. Analysis 2011;71:54-61.
[6] Beane SR, Davoudi Z, Savage M. Constraints on the universe as a numerical simulation. Eur Phys J A 2014;50.
[7] Kardashev NS. Transmission of Information by Extraterrestrial Civilisations [English Translation from Astrononicheskii]. Sov Astron A J 1964;8:217.
[8] Sagan C. Detectivity of Advanced Galactic Civilizations. Icarus 1973;19:350-352.
[9] Sagan C. Space Exploration as a Human Enterprise - Scientific Interest. Sci Public Aff 1973;29:30-33.
[10] Ashrafian H. Surgical Philosophy: Concepts of Modern Surgery Paralleled to Sun Tzu's 'Art of War'. CRC Press, 2015.
[11] Kaku M. Physics of the future : how science will shape human destiny and our daily lives by the year 2100. 1st edn New York: Doubleday, 2011.
[12] Weighill D, Jones P, Shah M, Ranjan P, Muchero W, Schmutz J *et al.* Pleiotropic and Epistatic Network-Based Discovery: Integrated Networks for Target Gene Discovery. Front Energy Res 2018;6.
[13] Villalonga B, Lyakh D, Boixo S, Neven H, Humble TS, Biswas R *et al.* Establishing the Quantum Supremacy Frontier with a 281 Pflop/s Simulation. arXiv:190500444 2019.
[14] Global_Challenges_Foundation. Global Challenges – Twelve risks that threaten human civilisation – The case for a new category of risks (https://api.globalchallenges.org/static/wp-content/uploads/12-Risks-with-infinite-impact.pdf). Oxford: Global Challenges Foundation, 2015.
[15] Haub C. How many people have ever lived on Earth?  Washington: Population Reference Bureau, 2011.
[16] Roser M, Ritchie H, Ortiz-Ospina E. World Population Growth (https://ourworldindata.org/world-population-growth/). OurWorldInData.org, 2019.
[17] UN. World Population Prospects - UN DESA (https://population.un.org/wpp/).  New York: United Nations, 2019.
[18] Cirkovic MM. The temporal aspect of the drake equation and SETI. Astrobiology 2004;4:225-231.